# Optimization of Transition Behaviors in a Two-Lane System

Yuming Dong[1], Xiaolu Jia[2], Daichi Yanagisawa[2,1,3], Akihito Nagahama[4], Katsuhiro Nishinari[2,1,3]

[1]Department of Aeronautics and Astronautics, School of Engineering, The University of Tokyo
7-3-1, Hongo, Bunkyo-ku, Tokyo, 113-8656, Japan
dong-yuming@g.ecc.u-tokyo.ac.jp
[2]Research Center for Advanced Science and Technology, The University of Tokyo
4-6-1, Komaba, Meguro-ku, Tokyo, 153-8904, Japan
[3]Mobility Innovation Collaborative Research Organization, The University of Tokyo
5-1-5, Kashiwanoha, Kashiwa-shi, Chiba, 277-8574, Japan
[4]Graduate School of Informatics and Engineering, The University of Electro-Communications
1-5-1, Chofugaoka, Chofu, Tokyo, 182-8585, Japan

**Corresponding author:**
E-mail address:dong-yuming@g.ecc.u-tokyo.ac.jp (Yuming Dong)

***Abstract.*** Transitions between two lanes often have a significant impact on various forms of road traffic. To address this problem, we have developed a two-lane asymmetric simple exclusion process model and two hypothetical traffic control strategies, to simulate a futuristic scenario where the timing and location of transitions between two lanes are highly controlled. Various scenarios were proposed to study the effectiveness of these control strategies. An optimized control strategy, whose parameters were determined through an optimization algorithm, is confirmed to effectively maximize the average traffic flow. Consequently, we may identify suitable road sections and the corresponding timings of transitions to resolve congestion in this model.

## 1. Introduction

Intelligent transportation systems (ITSs) and autonomous driving technologies are powerful tools to realize active traffic control that can mitigate issues; e.g., traffic congestion and pollutant emissions. For example, connected autonomous vehicles are expected to perform cooperative functionalities via the communication systems onboard [1]. In [2], traffic stability and flow in a highway-merging scenario was improved where the number of lanes decreased using a cooperative adaptive cruise control, in mixed traffic comprising cooperative adaptive cruise control vehicles and manually-driven vehicles. Further, in [3], cooperative merging was discussed from the viewpoint of optimal control. They confirmed that travel time and fuel consumption were improved by the proposed method: a central controller planned and controlled vehicle movements in certain control zones including the merging zone. In general, behaviors of vehicles may be completely changed by ITSs for optimizing the traffic in the future.

The control of transitions or lane changes could be a challenging and influential application of future ITSs. The transition behavior is sometimes considered "egocentric," because the vehicle involved often attempts to maximize its chance of accelerating without considering the traffic [4]. Based on previous studies, frequent transitions may cause severe Prisoner's Dilemmas, even influencing the traffic in large areas [5,6]. If vehicle behaviors in a large area are controlled by an ITS to avoid unnecessary transitions, the overall traffic flow may be expected to improve without degrading the benefits of individual drivers.

However, the simulation of ITSs on a relatively large scale has significant challenges [7-14]. Owing to the sheer complexity of this problem, a simple and effective model is required. The asymmetric simple exclusion process (ASEP) has successfully described particle transportation on a 1-D lattice, which was widely applied in areas such as traffic analysis, biological transport, and quantum mechanics [7–9]. If the particles in ASEP are only unidirectional, it is called a totally asymmetric simple exclusion process (TASEP) [10]. Two-lane ASEP is a noticeable development that is often considered an extension of the classic ASEP [11–13]. Two-lane ASEP allows more flexibilities for different internal states of particles, and/or simulates a more complicated traffic scenario. For example, allowing particles to jump between two lanes in ASEP may simulate the transition behaviors. In 2006, Pronina and Kolomeisky [14] performed a theoretical analysis on asymmetric coupling in a two-lane ASEP. For a full asymmetric scenario where the transition is only possible from one lane to the other, seven phases were determined for different inflow and outflow probabilities. Gupta and Dhiman [15] proposed a scenario where transition rates are determined by the configuration of the other lane. Nevertheless, the rates of transition between lanes are often determined before the start of the simulation, irrespective of the actual traffic.

On the other hand, the application of various traffic control strategies to control lane changes in ASEP has been validated. A simple feedback control based on the number of particles in the reservoir and the fluctuations of such a strategy were studied in [16]. The results of applying density feedback control strategy to a bottleneck situation in one-lane ASEP model is explored in [17]. Imai and Nishinari [10] developed a ASEP model for a forked road and discussed options for running an effective feedback control system. A feedback control based on density information has been proven to be more effective than other options, including average velocity and queue length. Most strategies aimed at improving the traffic with density feedback control (or similar methods) with a predetermined position for transitions and/or a fixed density threshold.

We proposed a two-lane TASEP model that regulates transitions by a novel control strategy. In our model, two concepts, transition position and density threshold, were proposed to simulate an ITS that can dynamically regulate transition behaviors between two lanes based on the density data. Such regulations are designed to allow more flexibility than traditional density feedback control, therefore differential evolution algorithm is applied to optimize the setup. This model has potential to further alleviate the problems of traffic congestion discussed in previous research mentioned above[10,16,17]. Besides, this scenario is likely to appear in the future because of the developments in ITSs and autonomous driving. We will mainly discuss the possibility for this hypothetical ITS to optimize the traffic by responding to the changing traffic conditions with different control strategies. The effectiveness and possible future applications of this hypothetical ITS be will also be studied.

The remainder of this paper is organized as follows. Section 2 introduces the working principle of the model, including the control strategy we applied and the differential evolution algorithm, which is necessary to determine the control strategy. In Section 3, we simplify the model and propose several different scenarios designed to simulate the model's response to the dynamic traffic pressure. The conclusions of these simulation results are presented in Section 4. Finally, the possible influence of applying the hypothetical ITS is discussed in Section 5.

## 2. Model
### 2.1 Overview
We developed a discrete-time TASEP model in MATLAB. The model is a two-lane lattice system with 240 sites on each lane. As a proof of concept, it is designed to be compatible with different scenarios, such as vehicles using the highway and pedestrians using the walkway. For each site, it is either empty or occupied. As shown in Fig. 1, particles are generated at the leftmost sites of the model (main entrances) and two side entrances located along lane 1, before moving to the rightmost sites for evacuation. The side entrances are designed to simulate the additional traffic flows from secondary roads or walkways.

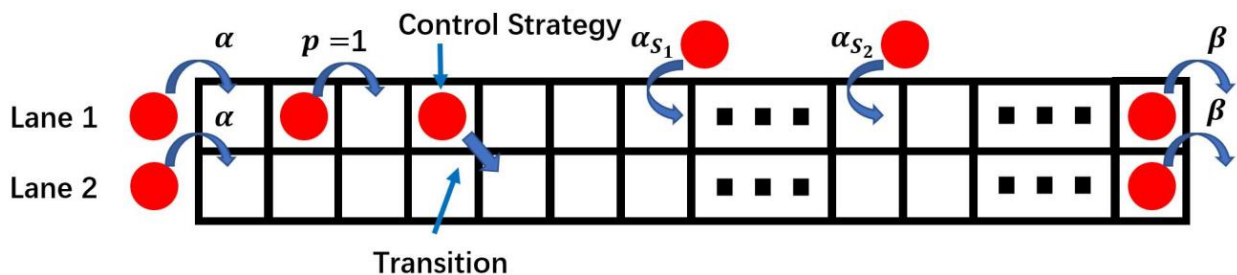

**Fig. 1** Overview of two-lane TASEP

The probability of particle generation (inflow probability) at the main entrances are denoted as $\alpha$ and the probability of particle evacuation (outflow probabilities) are denoted as $\beta$. The inflow probabilities of two side entrances are denoted as $\alpha_{S_1}$ and $\alpha_{S_2}$. Particles can hop to an empty site on the right with probability $p$. For simplicity, we assume that $p = 1$. The maximum speeds on each lane are denoted as $v_1$ (lane 1) and $v_2$ (lane 2). The maximum speed determines the number of sites that one particle can travel at most in one timestep [18]. Particles may perform a transition that allows them to jump to the other lane. The control strategy to determine the details of transitions is explained in Section 2.2.

As shown in Fig. 2, the two lanes are divided into 12 areas (areas 1-12), each with a length of 40 sites. Naturally, the sites on each lane can be denoted as integers from 1 to 240. These areas are predefined such that the model can track the particle density of each area without a significant increase in computation load. The particle density data are used in the control strategy described in Section 2.2.

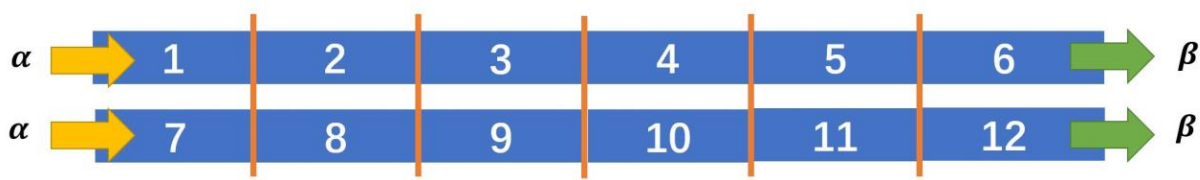

**Fig. 2** Two lanes are divided into 12 predefined areas. Main entrances and exits are marked with yellow arrows and green arrows, respectively.

### 2.2 Control Strategy

Allowing particles to perform transitions between two lanes in an ASEP model has been previously investigated [15-17]. The particle may hop to the other lane if the target site is empty, generally with a certain probability. However, we assume that transitions between lanes are completely controlled by a hypothetical ITS. Particularly, we developed a control strategy comprising two novel concepts: transition position and density threshold.

Transition Position (TP): we assume transitions can only be performed on certain sites referred to as transition positions. One TP occupies a single site and multiple TPs always occupy different sites. If a transition is ordered and the target site on the other lane is empty, the particle that is currently on the TP will be forced to perform a transition. The target site is located on the lower right for sites on lane 1 and on the upper right for sites on lane 2. If the target site is occupied, the particle will perform the normal movement on the current lane instead. The transition is completed in one timestep.

Density Threshold (DT): we assume that particles make transition decisions based on the particle density in the next area. Once the particle density exceeds the corresponding DT, transition orders will be given for all TPs in the previous area. For example, if two TPs are in area 2 and the DT for area 3 is 0.7, once the particle density of area 3 becomes larger than 0.7, particles reaching these two TPs are forced to perform transition (provided the target site is vacant). This is demonstrated in Fig. 3. This simple mechanism will only consider the density of area 3; therefore, area 9 can have a higher density than area 3 while transition orders are given.

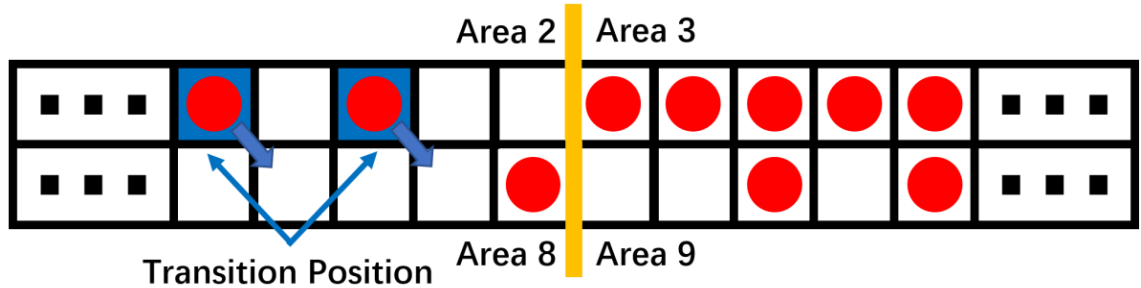

**Fig. 3** An example of performing transition while applying TP and DT

Transitions have priorities over normal particle movements. In other words, if the target site of a transition is also the target of a normal movement, the transition will always occur and particles performing the normal movement will consider this site as occupied. This represents the transitions' influence on traffic flow.

To maintain the main inflow and outflow uninfluenced by transitions, we assume that TPs cannot exist in the first and last areas, meaning there are 160 potential locations of TPs (site 41-200). In addition, because the particle density data of the next area are used to make transitions decisions, areas 6 and 12 would have no data to refer to. Irrespective of the number of TPs, the predetermined areas ensure that we only need to monitor the density of areas 3-6 and 9-12, which reduces the computation load. In reality, the density data may rely on devices with fixed locations and predetermined ranges, such as loop coil detectors.

A control strategy may be defined using a set of TPs and DTs. The average flow is used to evaluate the performance of a control strategy. For one simulation, the number of evacuated particles is obtained by averaging over 50 trials with 33000 timesteps for each. The first 3000 timesteps of each trail are the spin-up time and agents that complete the evacuation during this time are not recorded. Subsequently, the average flow is calculated by dividing the number of evacuated agents by 30000.

**2.3 Differential Evolution Algorithm**

If the number of TPs exceeds one, we cannot simply go through all combinations of TPs and DTs because of the sheer number of possibilities. We introduce the differential evolution algorithm as a solution. Proposed by Storn and Price in 1997 [19], this population-based stochastic optimization technique has been proven to be effective in several fields [19–21]. In terms of traffic flow, differential evolution algorithm has been applied to optimizing traffic signal timing at intersections [22,23], as well as solving short-term traffic flow prediction problem [24]. The flexibility of the algorithm [21] also allows application in vastly different scenarios, such as adjusting the geometrical parameters of obstacles that is designed for quicker emergency evacuation [25]. The working principle of the algorithm is briefly described in this section.

First, a population consisting of $N$ vectors need to be randomly generated within a $D$-dimensional space. The space is referred to as the search space, representing all possible solutions for the problem. In our case, the search space essentially covers every possible combination of TPs and DTs. Every $D$-dimensional vector of $N$ is called an individual. The individual may be converted to a set of TPs and DTs. Currently, we set that $N = 10$ and $D = 10$ (each lane has four DTs and one TPs). However, the value of $D$ will be changed to 5 in Section 3 to simplify the model and reduce the computation load. This change will not influence the following explanations.

The search space ranges of respective vector components are not the same. Because the differential evolution algorithm can only process continuous variables, DTs must be real numbers in [0,1]. The particle density from the model is discontinuous; therefore, DTs with similar values will have the same effect. For example, both 0.51 and 0.52 mean when 21 sites of the 40 sites are occupied, transition orders will be given. The corresponding DT is essentially 0.5. We may write the relationship as:

$$DT = \big(\text{floor}(x/0.025)\big) * 0.025, x \in [0,1],$$

where floor indicates the number is rounded down to the nearest integers toward zero.

On the other hand, the location of TPs is represented by real numbers in [0,1.6]. This relationship is as follows:

$$site = \big(\text{floor}(x/0.01)\big) + 40, x \in [0,1.6].$$

This equation essentially uses real numbers in [0,1.6] to represent sites 41-200. The differential evolution algorithm may remove unnecessary TPs through three mechanisms. First, it is possible to have $site = 40$, which is interpreted as "this TP does not exist" during the conversion process. Second, two TPs can have the same site value (E.g., 0.391 and 0.395 are all site 79). In that case, they are considered as a single TP. Third, $DT = 1$ means the corresponding TPs will not function. When will these happen is further discussed in Section 3.6. One example of converting a $D$-dimensional vector into TPs and DTs are given in Fig. 4.

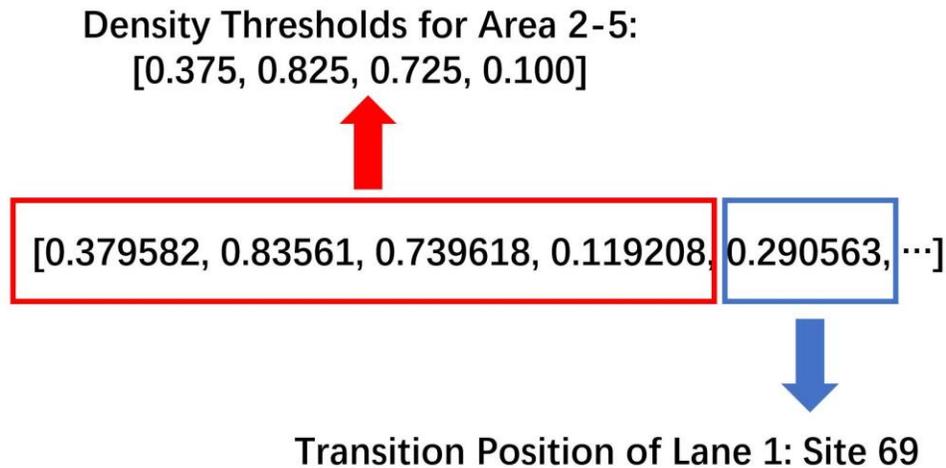

**Fig. 4** Converting a $D$-dimensional vector into TPs and DTs, which may be recognized by the TASEP model. If $D = 10$, there will be five additional numbers for representing the DTs and TPs for lane 2.

To initialize the population, the value of the $i$th individual on the $j$th dimension is randomly determined:
$$x_i^j = \text{rand}[0,1], j = 1,2,3,4,5,6,7,8$$
$$x_i^j = \text{rand}[0,1.6], j = 9,10,\ldots,D.$$

After the initialization, the fitness values of all individuals need to be calculated. Each individual will be converted to a set of TPs and DTs and imported to the TASEP model. Simulations will be performed to acquire the corresponding fitness values. The differential evolution algorithm will attempt to minimize the fitness value during the optimization process. Here, the fitness value is set to be the opposite of the average flow; that is, the differential evolution algorithm will attempt to maximize the average flow.

Next, a mutated individual is generated for each individual. Three random individuals ($x_{r_1}, x_{r_2},$ and $x_{r_3}$) need to be selected from the population, excluding the individual undergoing mutation. The mutated individual is defined as
$$M_i = x_{r_1} + F(x_{r_2} - x_{r_3}),$$
where $F$ is the differential weight. Here, $F$ is set to 0.85.

Each pair of mutated and original individuals will then undergo the crossover operation to generate a new trail individual $T_i$. For each dimension of $M_i$, a simple binomial crossover strategy is applied:
$$T_i^j = \begin{cases} M_i^j, & \text{if rand}[0,1] \leq P_{\text{CR}} \text{ or } j = j_{rand} \\ x_i^j, & \text{otherwise} \end{cases}, j = 1,2,\ldots,D,$$
where $P_{\text{CR}}$ is the crossover probability (set as 0.8 in our model) and $j_{\text{rand}}$ is a randomly selected dimension. Therefore, at least one dimension of the trial individual originates from the mutated individual [21]. All individuals will be checked if they have values that exceed the upper bounds or lower bounds. If so, these values will be reset to the upper or lower bounds. Through mutation and crossover, a new and potentially better population is generated from the old one. Differential evolution algorithm's mutation and crossover strategies are notably simpler than genetic algorithms, which may reduce the computation time.

Finally, the fitness value of every $T_i$ is again evaluated and are compared with the fitness values of $x_i$. If the fitness value of $T_i$ is not larger than that of $x_i$, $T_i$ will replace $x_i$. Otherwise, $T_i$ will be discarded and $x_i$ will remain unchanged. This process will ensure that the fitness value will always decrease after mutation and crossover. "Bad" mutations and crossovers will be discarded, while "good" ones are kept within the population.

One iteration comprises all the aforementioned steps. The differential evolution algorithm will repeat this process until 100 iterations have been performed and the best individual will be selected as the result. Fig. 5 provides a flow chart that describes the whole optimization process.

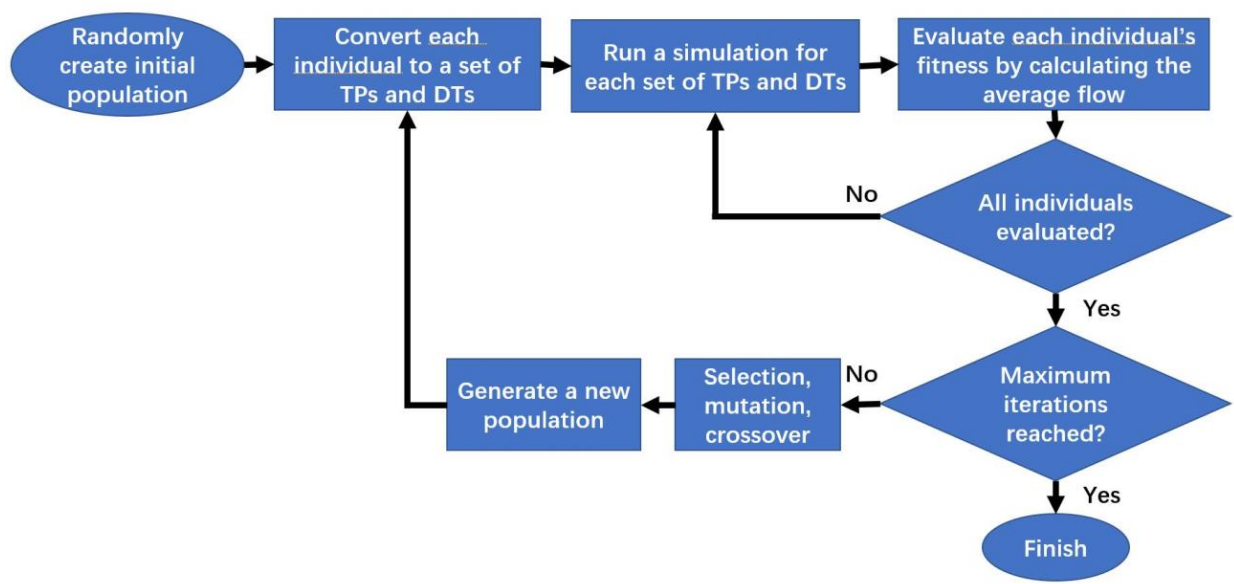

**Fig. 5** Block diagram of the optimization process.

## 3. Simulation and Results
### 3.1 Two Control Strategies
We considered two different control strategies: free transitions and optimized control strategies. "Free transitions" means for both lanes, the entire area between site 41 and site 200 is occupied by TPs and available to make transitions. DTs of four areas are set as 0.5.

TPs and DTs of the optimized control strategy are given by the differential evolution algorithm. Results from the differential evolution algorithm in this study are the averages of ten independent optimizations. Owing to the nature of our model, the differential evolution algorithm will likely generate some slightly different results during the ten independent optimizations; therefore, averaging is required. This is further explained in Section 3.3.

### 3.2 Simulation Setup
Normally, a scenario can be seen where two parts of a single path have different conditions. For example, a road may have a fast and slow lane, and an airport hallway may have a moving and a normal walkway. Therefore, the setup of the model is designed based on the following hypothetic scenario.

Two lanes are set to have different conditions. Lane 1 has a lower speed limit and must cope with the extra traffic coming from side entrances. Lane 2 has a higher speed limit, although transitions from lane 1 may interrupt the high-speed flow. Ideally, a good control strategy should maintain a balance between the two lanes—redirecting the extra pressure of one lane to the other, while limiting the influence of transitions on the high-speed flow on lane 2. Moreover, the hypothetical ITS should respond to changing inflows and outflows by preparing multiple control strategies. Fig. 6 provides an illustrative example where the inflows of main entrances are increasing and positions of inflows of side entrances are changing, as the time changes from T1 to T2. Ideally, the control strategy may respond to the changes by shifting to another set of TPs and DTs. This possibility is further discussed in Section 3.7.

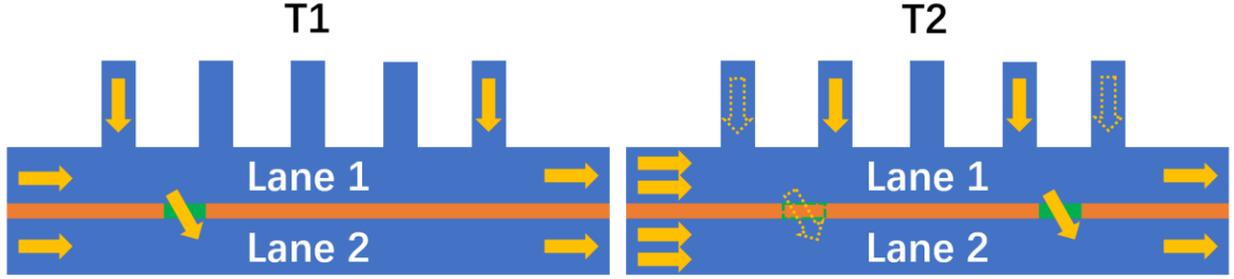

**Fig. 6** A hypothetical scenario where different control strategies can be applied to different traffic conditions.

In this setup, we assume that $v_1 = 1$ site/timestep and $v_2 = 2$ sites/timestep. If the site ahead is occupied, the particle will stop completely. If there is only one vacant site ahead, the speed of the particle will decrease to 1 site/timestep. For simplicity, there is no acceleration or deceleration time. Owing to the increased maximum speed, we assume that in lane 2, particles can also evacuate through the second site from the right. This asymmetric setting is critical to our following results. If $v_1 = v_2$, regulating transitions has a limited effect.

We considered 9 possible scenarios with different combinations of $\alpha$, $\beta$, $\alpha_{S_1}$, and $\alpha_{S_2}$. The inflows of side entrances represent the dynamic traffic pressure from secondary roads. The particles generated at the side entrances will try to join traffic flow of lane 1 if possible, similar to how real-life traffic from secondary roads merge with the traffic on the main road. As summarized in Table 1 and Fig. 7, three sets of side entrance positions and $\alpha_{S_i}$ are considered. They are named SE1-SE3 (SE means side entrances).

| Name | SE1 | SE2 | SE3 |
|---|---|---|---|
| Positions of Side Entrances | 121, n/a | 81, 161 | 41, 201 |
| $\alpha_{S_1}$ and $\alpha_{S_2}$ | 0.2, n/a | 0.1, 0.1 | 0.1, 0.1 |

**Table 1** Three sets of side entrances and the corresponding $\alpha_{S_1}$ and $\alpha_{S_2}$. For SE1, only one side entrance exists, therefore the position of the second side entrance and $\alpha_{S_2}$ are marked as "n/a".

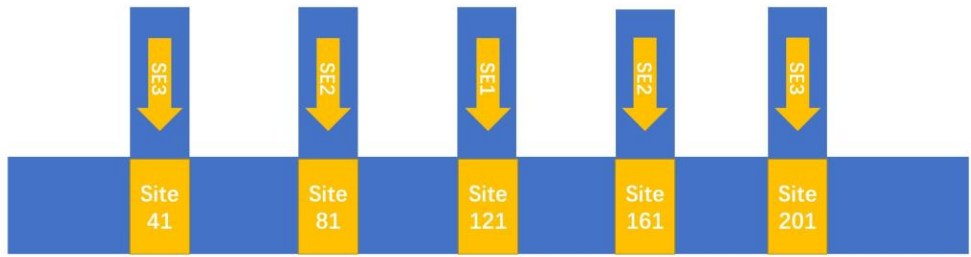

**Fig. 7** Lane 1 and positions of side entrances. Note that the site sizes are purely illustrative.

Additionally, as shown in Table 2, we have three sets (IO1-IO3, IO refers to inflow and outflow) of inflow and outflow probabilities. Because the hypothetical control strategy is designed to optimize traffic flow when under pressure, $\alpha$ is always larger than $\beta$.

| Name | IO1 | IO2 | IO3 |
|------|-----|-----|-----|
| $\alpha$ | 0.6 | 0.8 | 1 |
| $\beta$ | 0.4 | 0.6 | 0.8 |

**Table 2** Three sets of inflow and outflow probabilities

Therefore, nine different combinations are present, representing the dynamic traffic pressure that must be withstood by the hypothetical ITS. We applied the differential evolution algorithm to determine an optimized control strategy in each case and compare the results with those from the free transitions control strategy.

### 3.3 Simplification of Transition Direction

As seen in Section 3.2, the extra pressure from side entrances likely makes two-way transitions detrimental to the overall traffic flow. The transitions should balance the burden of two lanes instead of increasing the difference. We can confirm this by simulating a number of test scenarios where SE2 is applied. As summarized in Table 3, transitions are either only performed from lane 1 to lane 2 (one-way) or performed in both directions (two-way).

| Scenario | | Average Flow (Number of Evacuated Agents/Timestep) | |
|---|---|---|---|
| | | Free Transitions | Optimized |
| IO1 | One-Way | 0.7475 | 0.7594 |
| | Two-Way | 0.7485 | 0.7593 |
| IO2 | One-Way | 0.8615 | 0.9466 |
| | Two-Way | 0.8436 | 0.9466 |
| IO3 | One-Way | 0.9097 | 1.0414 |
| | Two-Way | 0.8908 | 1.0405 |

**Table 3** Changes in the average flow and optimized control strategy if transitions cannot be performed in both directions. In all "One-Way" scenarios, transitions may only be performed from lane 1 to lane 2. In all scenarios, we apply SE2 defined in Table 1.

From the results, transitions from lane 2 to lane 1 have highly limited influence on the average flow. If transitions are only allowed to be performed from lane 1 to lane 2, the number of variables will be halved ($D = 5$ instead of 10) and the computation load will be significantly reduced. In additional, the hypothetical ITS will only need to monitor the densities of four areas instead of eight. Therefore, we only allow transitions from lane 1 to lane 2 in remainder simulations of the paper. This will largely simplify the model and improve the computation speed. From a relatively realistic perspective, this will also reduce the need of traffic density information for the hypothetical ITS. After this simplification, a single optimization process takes around one hour on an Intel® i7-11800H @ 2.30GHz processor when running on eight cores.

**3.4 Influence of Inflow and Outflow**

The simulation results of nine scenarios mentioned in Section 3.2 are listed in Table 4.

| Scenario | | Average Flow (Number of Evacuated Agents/Timestep) | | TP | DT |
|---|---|---|---|---|---|
| | | Free Transitions | Optimized | | |
| IO1 | SE1 | 0.7474 | 0.7592 | 200 | 0.600 |
| | SE2 | 0.7475 | 0.7594 | 200 | 0.500 |
| | SE3 | 0.7536 | 0.7593 | 200 | 0.600 |
| IO2 | SE1 | 0.8598 | 0.9477 | 109 | 0.400 |
| | SE2 | 0.8615 | 0.9466 | 69 | 0.375 |
| | SE3 | 0.8827 | 0.9303 | 187 | 0.450 |
| IO3 | SE1 | 0.9064 | 1.0437 | 78 | 0.375 |
| | SE2 | 0.9097 | 1.0414 | 67 | 0.450 |
| | SE3 | 0.9381 | 1.0239 | 78 | 0.375 |

**Table 4** Results from free transitions and optimized control strategy. TPs and DTs of the optimized control strategy are also shown. Note that TPs and DTs are from optimized control strategies, and already converted to numbers recognizable by the two-lane TASEP model. Similar approach has been applied throughout the paper.

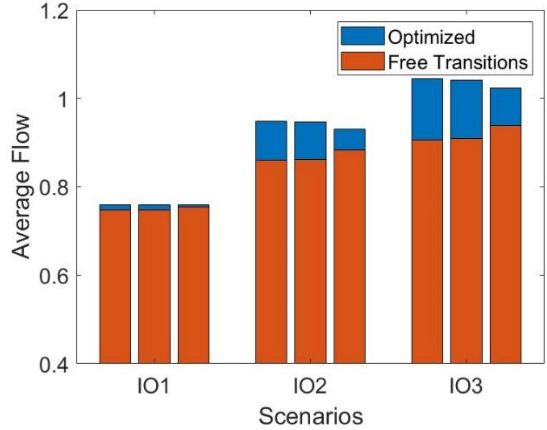

**Fig. 8** Average flows of nine scenarios in Table 4. For each IO, the three bars denote average flows corresponding to SE1, SE2, and SE3 (from left to right).

Generally, optimized control strategies have provided better results than free transition control strategy. Optimized control strategies better adapt to the large inflow probability, whereas the inflexibility of the free transitions control strategy is disadvantageous. This disadvantage, however, is less obvious when the outflow probability is low, which can be clearly observed in results of IO1 in Fig. 8. The exact values of optimized control strategies' TPs and DTs depend on $\alpha$, $\beta$, $\alpha_{S_1}$, and $\alpha_{S_2}$. The values are mostly unique; therefore, there is no "one-size-fits-all" control strategy. However, in Table 4, we may see that in five out of nine scenarios, TPs are located within the right half of area 2 (site 60-80). We may safely assume that performing transitions in this area will improve the traffic flow. For IO1, TPs are located near the rightmost sites of area 5, though the insignificant difference between optimized control strategy and free transitions control strategy cannot prove the effectiveness of such TPs.

Although the side entrances also have a significant influence on the average flow, as demonstrated in Fig. 8, optimized control strategies have demonstrated a significant advantage for SE1 and SE2. For SE3, the performances of optimized control strategies are generally closer to those of the free transitions control strategy. Even with the same inflow probability ($2\alpha + \alpha_{S_1} + \alpha_{S_2}$), the extra traffic flow coming in from side entrances away from the center will significantly influence average flow. This effect is evident for SE3.

Two sets of raw results (not full) are listed in Table 5 as examples to discuss the acquisition of the optimized control strategy from raw results and why multiple optimized control strategies have similar performances. If $\alpha = 0.6, \beta = 0.4$ (IO1), the average flows are stable even when vastly different control strategies are applied. This indicates the existence of a "plateau," i.e., no matter how the differential evolution algorithm adjusts the values of TPs and DTs, the average flows are still limited by other factors such as $\beta$. Discussing the errors of the values of TPs and DTs or averaging over them are meaningless. Therefore, only one set of TP and DT is provided in Table 5. Conversely, if $\alpha = 0.8, \beta = 0.6$ (IO2), the differential evolution algorithm can provide stable average flows and a single optimized control strategy. This indicated the existence of a "peak," i.e., if a proper control strategy is applied, the maximum average flow may be reached.

As mentioned in Section 2.3, DTs with values in [0.375,0.400) mean if 16 out of 40 sites are occupied in the next area, transition orders will be given. Therefore DTs with values in [0.375,0.400) are all written as 0.375 in Table 4 and 5.

| IO1-SE2 | | | IO2-SE2 | | |
|---|---|---|---|---|---|
| Average Flow | TP | Corresponding DT | Average Flow | TP | DT |
| 0.7596 | 200 | 0.525 | 0.9466 | 69 | 0.375 |
| 0.7595 | 100 | 0.275 | 0.9468 | 69 | 0.375 |
| 0.7595 | 103 | 0.150 | 0.9466 | 69 | 0.375 |
| 0.7595 | 133 | 0.650 | 0.9465 | 69 | 0.375 |
| 0.7595 | 92 | 0.450 | 0.9467 | 69 | 0.375 |

**Table 5** Optimized control strategies and the corresponding average flows when IO1-SE2 and IO2-SE2 are applied.

Finally, we can confirm that the differential evolution algorithm provided the best control strategy by undergoing all possible combinations of TP and DT. This process is time-consuming; therefore, we only perform this check on IO1-SE2 and IO2-SE2. The results are shown in Fig. 9. We may see the "peak" which represents the best combination of TP and DT in the right plot. The left plot, however, contains a "plateau" that indicates a limited maximum average flow. This is why the differential evolution algorithm has often provided more than one optimized control strategy.

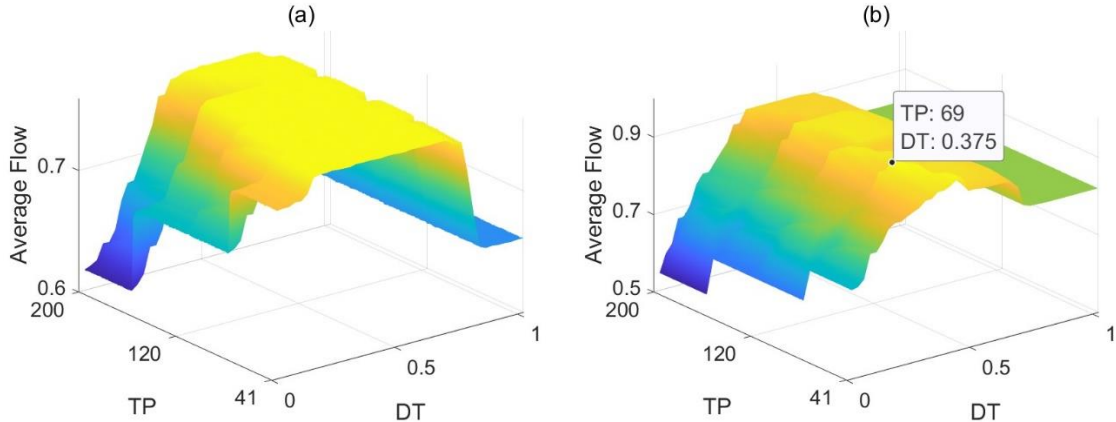

**Fig. 9** Average flows of all possible combinations of TP and DT when IO1-SE2 (a) and IO2-SE2 (b) are applied. The "peak" of IO2-SE2 is marked.

### 3.5 Stochastic Variation

When the values of TP and DT are fixed, the average flow still fluctuates within a small range because of random factors such as inflow and outflow probabilities. Therefore, for each result, the average flow is calculated from 50 simulations. We discussed the errors by acquiring 400 results from a test scenario where IO2-SE2 is applied. We assume that for this test scenario, TP is located on site 120 and the corresponding DT is 0.5. As shown in Fig. 10, the random error of the model is insignificant.

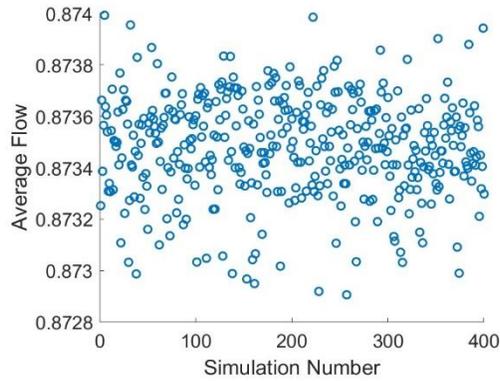

**Fig. 10** 400 simulation results from the test scenario. The maximum is 0.8740 and the minimum is 0.8729.

Another source of error is the spin-up time. We tested whether 3000 timesteps is long enough for the model to become relatively stable by simulating the same test scenario 1000 times. As shown in Fig. 11, the flow rate will become relatively stable after 1000 timesteps. The flow rate is calculated by averaging over the flow rate at that timestep in each of the 1000 simulations.

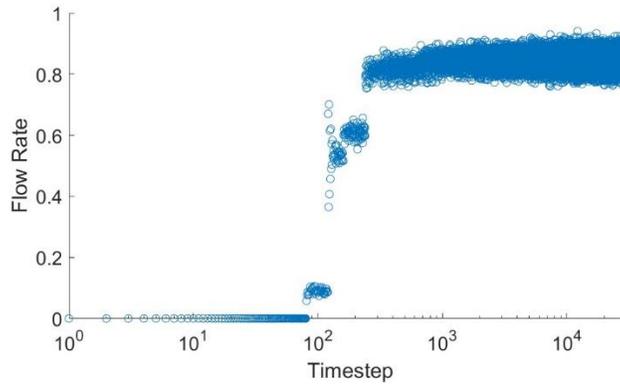

**Fig. 11** Flow rates of the 33000 timesteps. The flow rates are the average of 1000 simulations of the test scenario.

### 3.6 Increase in the Number of TPs

Naturally, we may speculate that an increase in the number of TPs in the optimized control strategy influences the average flow. However, when the number of TPs is increased to two, such influence is small. IO1-SE2 and IO2-SE2 are chosen again for comparison. The results are listed in Table 6.

| Scenario | | Average Flow | | TP | DT |
|---|---|---|---|---|---|
| | | Free Transitions | Optimized | | |
| IO1-SE2 | 1 TP | 0.7475 | 0.7594 | 200 | 0.500 |
| | 2 TPs | | 0.7593 | 113, 200 | 0.675, 0.650 |
| IO2-SE2 | 1 TP | 0.8615 | 0.9466 | 69 | 0.375 |
| | 2 TPs | | 0.9467 | 69, 74 | 0.375, 0.375 |

**Table 6** Results from the free transitions and optimized control strategies with two TPs. The TPs and DTs of the optimized control strategy are also shown.

Compared with the results in Section 3.3, the difference in average flow is not significant and the corresponding DTs are generally larger. While one TP may be enough for reaching the highest average flow, the transitions may also be even out between two TPs to achieve a similar effect. Therefore, one or two TPs are sufficient to optimize the traffic if proper control strategies are applied. We may safely assume that if three or more TPs are applied, the average flow will not increase and the differential evolution algorithm will often remove excessive TPs through mechanisms mentioned in Section 2.3. Moreover, this indicates that the hypothetical ITS may significantly influence the traffic with relatively limited orders, provided it can estimate the overall traffic condition and react quickly.

### 3.7 Effect of the Optimized Control Strategy with Respect to the Outflow Probability

We explored when control strategies become inefficient in optimizing the traffic flow. We simulated IO2-SE2 ($\alpha = 0.8, \beta = 0.6$) and gradually decreased $\beta$ to 0.2. A sampling interval of 0.05 was chosen because of the limitations of the computing power. The results are shown in Fig. 12.

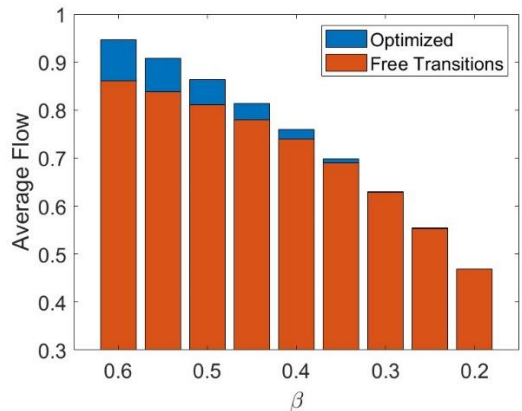

**Fig. 12** Changes in the average flow when $\beta$ gradually decreases from 0.6 to 0.2

As seen, when $\beta_1$ and $\beta_2$ decreased, the advantage of the optimized control strategy will finally disappear. Despite the absence of a sharp turning point, the average flow of the optimized control strategy decreased faster after $\beta_1$ and $\beta_2$ reached 0.55, which is also the point where the optimized control strategy demonstrated the most significant advantage. This indicates that if $\beta_1$ and $\beta_2$ suddenly start decreasing for reasons such as traffic accidents, the hypothetical ITS may largely mitigate the negative influence by applying a different control strategy until it is overwhelmed by finally increasing traffic pressure. The hypothetical ITS should be able to quickly notice the decrease in the outflow.

## 4. Conclusion

In this study, we developed a two-lane TASEP model to analyze the effectiveness of controlling when and where transitions between two lanes can be performed. This optimized control strategy is defined using two novel concepts, namely TPs and DTs. The exact values of TPs and DTs were provided using a differential evolution algorithm.

We confirmed the validity of optimizing traffic by providing different control strategies for different traffic conditions, provided data such as inflow, outflow, and density can be quickly estimated. Nine scenarios with different inflow and outflow probabilities were tested to simulate different traffic conditions. Based on the results, one or several optimized control strategies can be applied to maximize traffic flow. In addition, we explored the point when the optimized control strategy ceases to be effective.

The optimized control strategy includes marking one or two sites as TPs and one of two corresponding DTs. With sufficient data, we may identify key areas hat is most suitable for performing transitions and the necessary actions to optimize the traffic in each scenario. In the model, the key area is the right half of area 2. The possible applications of this conclusion and future research directions will be discussed in Section 5.

## 5. Discussion

In the future, traffic congestion will likely be significantly reduced as autonomous vehicles and ITSs provide more methods to control and optimize traffic. Although this study is still mostly a proof of concept, we may expect future ITSs to apply our findings and optimize the traffic through careful regulations of transition behaviors. The idea of a control strategy may seem futuristic, although similar scenarios can still be found in real life. In large airports, moving walkways are often placed one after another. If the moving walkway and the hallway are considered two separate "lanes," the space between two moving walkways is similar to our notion of TP, which is a fixed location where people may perform a "transition."

The optimization process is relatively time-consuming; therefore, real-time responses could be a challenge. One possible solution is to simulate multiple possible scenarios beforehand and prepare corresponding control strategies. For example, if each area is assumed to be 1 km long and 1 site/timestep means 40 km/h, then 1 timestep is 2.25 s. To regulate the transition behaviors, the hypothetical ITS may apply the most suitable optimized control strategy from its database of 400 timesteps (15 min), before reevaluating the situation and applying a new strategy. With an increase in computation power, pre-calculated control strategies may be partially or even entirely replaced by those generated based on site information. Theoretically, these regulations on transition behaviors will improve the road traffic condition by minimizing their influence on the traffic flow. The system might even gradually improve itself by collecting traffic data and assessing the influence of its control strategies with time. On the other hand, given that key areas can already be located, we may limit transitions to this area with current technology (by putting up fences between lanes, for example). Considering the psychological influence on drivers, whether this is a viable approach in real life is nevertheless questionable.

Future ITSs are expected to be faced with more complicated traffic conditions than our TASEP model; therefore, a more detailed TASEP model will be required in the future. Further, the ITS should have more tools and resources at its disposal. We may easily imagine a scenario where inflow probabilities of side entrances can be adjusted by the ITS. For example, if another parallel lane connects to lane 1 by the side entrances, the system may determine the timings of making right turns to achieve this. When area 3 becomes congested, the system can order transitions in area 2, while redirecting traffic away from the leftmost entrance to further reduce the pressure. Our proposed control strategy is designed to only have one threshold, although future ITSs will presumably have more carefully designed options. For example, once the transition order is triggered, it will continue until a second threshold is reached, thereby controlling the density more accurately. The density on the other lane may also be considered during the decision process.

In addition to developing more complicated ASEP models, their abstract nature means a more detailed model needs to be established in the future to further explore the control of transitions behaviors. For example, the difference between maximum speeds may be closer to reality in a more detailed model. Moreover, the viability of applying almost absolute control on transition behaviors in real traffic is worth discussing because, in the future, we cannot simply assume that every human driver will follow orders from the ITS. This "disobedience" should be simulated for accuracy. A much more detailed model and a completely different approach for finding out the optimal TPs (narrow the possible locations down to a small area by using our current model, for example) will be required for future study. Finally, it might be possible to simulate pedestrian movements with a similar model. In this case, lane 2 will be a series of moving walkways, and TPs will be the space between two moving walkways. Although that does leave out the question about how to give pedestrian transition orders, we speculate that it may still provide insights about reducing the evacuation time in certain occasions like a heavily used underground walkway system.

**Acknowledgment**

This work was supported by JSPS KAKENHI grant numbers JP21H01570, 21H01352, and 21K14377, and JST-Mirai Program Grant Number JPMJMI20D1, Japan.

**Competing interest**

We declare that we have no conflict of interest.